\documentstyle[12pt,epsfig]{article}

\large
\newcommand{\be}{\begin{eqnarray}}
\newcommand{\ee}{\end{eqnarray}}
\newcommand{\bea}{\left (\begin{array}{cc}}
\newcommand{\eea}{\right )\end{array}}

\newcommand{\Dirac}{\rlap {\hspace{-0.5mm} \slash} D}
\begin{document}
\setlength{\baselineskip}{17pt}
\pagestyle{empty}
\vfill
\eject
\begin{flushright}
SUNY-NTG-99/8
\end{flushright}

\vskip 2.0cm
\centerline{\Large \bf The Spectral Density of the QCD Dirac Operator} 
\vskip 0.4cm
\centerline{\Large \bf and Patterns of Chiral Symmetry Breaking}

\vskip 1.2cm
\centerline{D. Toublan and J.J.M. Verbaarschot}
\vskip 0.2cm
\centerline{Department of Physics and Astronomy, SUNY, 
Stony Brook, New York 11794}
\vskip 1.5cm

\centerline{\bf Abstract}
We study the spectrum of the QCD Dirac operator 
for two colors with fermions in the fundamental representation 
and for two or more colors with  adjoint fermions.
For $N_f$ flavors, the chiral flavor symmetry of these theories 
is spontaneously broken according
to $SU(2N_f)\rightarrow Sp(2N_f)$ and $SU(N_f)\rightarrow O(N_f)$, 
respectively, rather than the
symmetry breaking pattern $SU(N_f) \times SU(N_f) \rightarrow SU(N_f)$ 
for QCD with three or more colors and fundamental fermions. 
In this  paper we study the Dirac spectrum for the first two
symmetry breaking patterns. Following previous work for the third 
case we find the Dirac spectrum
in the domain  $\lambda \ll \Lambda_{\rm QCD}$
by means of partially quenched chiral perturbation theory. 
In particular, 
this result allows us to calculate the slope of the Dirac spectrum at
$\lambda = 0$. We also show that for $\lambda \ll 1/L^2 \Lambda_{QCD}$
(with $L$ the linear size of the system)
the Dirac spectrum is given by a chiral Random Matrix Theory with
the symmetries of the Dirac operator.


\vfill
\noindent

\eject
\pagestyle{plain}

\vskip1.5cm
\noindent
{\bf 1. Introduction}
\vskip 0.5cm
Perhaps the most important reason for the current interest in the 
infrared sector of the Euclidean Dirac
spectrum is its relation with the chiral condensate by means of the 
Banks-Casher formula \cite{BC}. In this way the properties 
of the smallest eigenvalues 
of the Dirac operator may be related to the mechanisms of chiral 
symmetry breaking and to the nature of the chiral phase transition 
as observed in 
lattice QCD simulations \cite{DeTar,Smilref}. 
However, this is not the only reason for our interest
in this part of the Dirac spectrum. What may be
more important ultimately is that it is one of the few examples where 
exact nonperturbative analytical results can be derived from QCD.

One of the main objectives of this paper is to study  the 
QCD Dirac spectrum in the domain $\lambda \ll \Lambda_{\rm QCD}$.
In particular we calculate the slope of the Dirac spectrum near
$\lambda = 0$. 
This problem  was first addressed in a paper by Smilga and Stern
\cite{SmilgaStern}. They 
studied the slope of the Dirac spectrum for the case of three or
more colors with fermions in the fundamental representation. 
In this article we extend previous
results \cite{SmilgaStern,OTV,DOTV}
to the case of QCD with two colors and fundamental fermions and
QCD with adjoint fermions. Before discussing our strategy let us first 
summarize the approach followed by Smilga and Stern. They studied the
scalar susceptibility which, in terms of the spectral density 
averaged over the Euclidean QCD action,
\be
\rho(\lambda) = \sum_k \langle \delta(\lambda-\lambda_k) \rangle,
\ee
is given by
\be
K^{ab} = -\delta^{ab} \int_0^\infty \frac {\rho(\lambda) (m^2 - \lambda^2)}
{(m^2 + \lambda^2)^2} d\lambda.
\label{scalar}
\ee
Their idea was to derive  $K^{ab}$ from chiral perturbation
theory and invert this relation to obtain information on the Dirac
spectrum. Because of the implicit mass dependence of $\rho(\lambda)$ via 
the fermion determinant in the statistical average over gauge field 
configurations such inversion is not possible in general. They found that
\be
K^{ab} \sim \delta^{ab}\frac{N_f^2 -4}{N_f} \log(m/\Lambda),
\ee
where $\Lambda$ is the momentum cutoff of the one-loop integral. Their crucial
observation was that the contribution
to the integral (\ref{scalar}) from the chiral logarithms in $\rho(0)$ 
cancels because
\be
 \int_0^\infty \frac {\rho(0) (m^2 - \lambda^2)}
{(m^2 + \lambda^2)^2} d\lambda =0.
\ee
Barring any more exotic $\lambda$ dependence of  $\rho(\lambda)$ 
such as for example $\sqrt {m\lambda} \log m$, the logarithmic mass
dependence of the scalar susceptibility results entirely from the
linear term in $\rho (\lambda)$.

However, there is another approach to extract the Dirac spectrum directly
from the low energy effective partition function which does not
suffer from the above mentioned ambiguities in the mass dependence
\cite{Vplb,OTV,DOTV}. The starting point of
this approach is the so called partially quenched partition function 
defined as \cite{Morel,pqChPT,GolLeung}
\be
Z^{\rm pq}(m_v,J)  ~=~ \int\! [dA] 
~\frac{\det(\Dirac +m_v+J)}{\det(\Dirac +m_v)}\prod_{f=1}^{N_{f}}
\det(\Dirac + m_s) ~e^{-S_{YM}[A]} ~,
\label{pqQCDpf}
\ee
where $\Dirac$ is the Euclidean Dirac operator and $S_{YM}[A]$ is the
Euclidean Yang-Mills action.
The valence quark mass dependence of the chiral condensate is then given
by \cite{Christ,Vplb,Trento}
\be
\Sigma(m_v) = \frac 1V\langle \sum_k \frac 1{m_v +i\lambda_k} \rangle  =
\frac 1V \left . \partial_J \right |_{J=0} \log Z^{\rm pq} (m_v, J).
\ee
The spectral density then follows from the discontinuity across the imaginary
axis
\be
\rho(\lambda)/V = \frac 1{2\pi} (\Sigma (i\lambda +\epsilon) - 
\Sigma( i\lambda - \epsilon)).
\ee

The valence quark mass dependence of the partition function 
(\ref{pqQCDpf}) can be obtained from its corresponding low-energy effective
partition function \cite{qChPT,pqChPT,GolLeung,Sharpe}
which is based on the specific
pattern of chiral symmetry breaking.
In this way it is possible to extract exact analytical information on
the QCD Dirac spectrum in the domain near zero virtuality. This approach
was first followed in \cite{OTV,DOTV} for three or more colors with 
fundamental fermions  where in addition to extending the result of Smilga and
Stern we were able to calculate the spectral density of the Dirac operator
on the scale of the smallest eigenvalues.
In this paper we apply this approach to QCD with two 
colors with fermions in the fundamental representation and 
QCD with adjoint fermions.

A much simpler theory with the global symmetries of the QCD partition function
is chiral Random Matrix Theory (chRMT) \cite{SVR,V}. 
In this theory the matrix elements
of the Dirac operator are replaced by Gaussian distributed random variables.
It turns out that this does not affect the pattern of chiral symmetry breaking.
The separation into three different symmetry classes can be related
naturally to the Dyson index of the Dirac operator denoted by the 
symbol $\beta$. Its value is $\beta =1$ for QCD with two colors and fermions
in the
fundamental representation, it is $\beta = 2$ for QCD with three or more
colors and fermions in the fundamental representation, 
and it is $\beta =4$ for QCD
with adjoint fermions. The matrix elements of the Dirac matrix of the 
corresponding chRMT are real, complex and quaternion real, 
respectively \cite{V,SmilV}.
Within  chRMT it is possible
to obtain exact analytical expressions for the spectral density and 
all spectral correlation functions. As we will see in the next paragraph
it is possible to establish a domain where the spectral correlations of 
the QCD Dirac operator are given by chRMT. Alternatively,
the $existence$ of such domain can be understood by means of universality
arguments 
\cite{brezin,Damgaard,Sener1,GWu,Seneru,Dampart,DAm3dmass,Christiansen}.

We can distinguish three important scales in the QCD Dirac spectrum. They are
most conveniently  discussed in terms of the the valence quark mass.
The first scale is  the smallest nonzero eigenvalue
of the Dirac operator which is of the order of the average level spacing and
is given by
\be
\lambda_{\rm min} \sim \frac \pi{\Sigma_0 V},
\label{lmin}
\ee
where $\Sigma_0$ is the chiral condensate in the chiral limit and $V=L^4$
is the volume of space-time.
The second scale corresponds to the valence quark mass for which the
Compton wave length of the associated Goldstone
boson  is equal to the size of
the box \cite{Vplb,Osbornprl,OTV},
\be
\frac {m_v \Sigma_0 }{F^2} \sim \frac 1{L^2},
\ee
As was argued in \cite{Vplb,Osbornprl,OTV,DOTV} 
the valence quark mass dependence of the 
chiral condensate is described by chRMT for 
\be
m_v \ll m_c =\frac {F^2}{\Sigma_0 L^2}.
\label{mc}
\ee
The spectral density of the Dirac operator, given by the discontinuity 
of $\Sigma(m_v)$, is therefore described by chRMT as well
in this domain. 
These  results have been confirmed by numerous lattice QCD simulations
\cite{Halasz,Vplb,Tilo,many,Ma,Guhr-Wilke,Tilomass,andystudent}
\cite{markum,Dam3d,DamSU3,TiloSU3,Berg,Hipschwing,Heller,Kiskis},
instanton liquid simulations \cite{Vinst,Osbornprl,Osbornnpb} and
other calculations \cite{Simons,Takahashi}.
For valence quark masses below this scale the kinetic term
in the effective Lagrangian can be ignored. In \cite{OTV,DOTV} it was
shown that in this regime the valence quark mass dependence
that follows  from partially quenched chiral perturbation theory
coincides with the result from chRMT. 
The dimensionless ratio of the scales (\ref{lmin}) and (\ref{mc}) 
\be
\frac{m_c}{\lambda_{\rm min}} \sim \frac{F^2 L^2}\pi
\ee
represents  the  approximate number of eigenvalues    
that are described by chRMT.
A scale analogous to $m_c$ occurs in mesoscopic physics where it is known
as the Thouless energy \cite{Altshuler,spreading}. 

Based on the scales 
\be
\lambda_{\rm min}  \ll m_c \ll \Lambda_{QCD},
\ee
we thus can distinguish four  different domains. 
With names borrowed from mesoscopic physics, these scales separate
the quantum domain, the ergodic domain, the diffusive domain and the 
ballistic domain. The valence quark masses can be interpreted in terms of
a conjugate time. Then  $\hbar/m_c$ is the
time scale for which a particle diffuses over the length 
of the system with diffusion constant given by $F^2/\Sigma_0$
\cite{Polonyi,Osbornprl,Janik}.
For more discussion and additional references on impurity scattering 
in disordered systems and the Thouless energy we refer to
\cite{Efetov,HDgang,Montambaux}.

In the domain of validity for chRMT it is natural to study the spectral
density in units of the average level spacing.  
The corresponding microscopic spectral density is defined by \cite{SVR}
\be
\rho_s(u) = \lim_{V\rightarrow \infty} 
\frac 1{\Sigma V} \rho(\frac {u}{\Sigma V} ).
\ee
In this paper, we will show by means of a perturbative calculation
that in the ergodic domain 
the microscopic spectral density that follows from
partially quenched chiral perturbation theory coincides with the result
obtained from the chiral Random Matrix Theory in the same universality
class. For smaller values of $m_v$ the super-integrations have to be
performed exactly. This has been achieved \cite{OTV,DOTV}
for the case $\beta = 2$ which is mathematically much simpler than the
cases $\beta=1$ and $\beta=4$ \cite{Efetov,Wegner,VWZrep}
and complete agreement with chRMT was found.
It would be interesting to perform a similar calculation for the other two
cases as well.

As second important result of this paper is the derivation of the
slope of the Dirac spectrum for QCD with two colors and fundamental 
fermions and for QCD with adjoint fermions. 
We will derive this result in two different
ways. First, according to the method of Smilga and Stern, and second by means
of the partially quenched QCD partition function. We find that both methods
agree thereby excluding a more exotic mass dependence of the spectral
density which would be allowed by the Smilga-Stern approach.
Our result for the slope of the Dirac spectrum 
for $N_f$ massless fermions can be summarized as
\be
\rho'(\lambda = 0) = \frac {\Sigma^2_0}{16\pi^2 F^4} 
\frac {(N_f-2)(N_f +\beta)}{\beta N_f},
\ee
where $\Sigma_0 $ is the chiral condensate, $F$ is the pion decay constant and
$\beta $ is the Dyson index of the Dirac operator. 

In the next two sections we study the symmetry breaking patterns
for $\beta=1$ and $\beta=4$. The slope of the Dirac spectrum for these
two cases is calculated from standard ChPT in section 4.
Partially quenched chiral perturbation theory based on supersymmetry
is constructed in section 5. Within these effective theories we calculate
the valence quark mass dependence of the chiral condensate. The computations
for two colors with fundamental fermions and adjoint fermions are presented
separately in sections 6 and 7, respectively. Some interesting limiting cases
and compact expressions for all $\beta$ are given in section 8. In section
9 we show that our results agree with chRMT in the ergodic domain. 
We also find the Dirac spectral density for
larger eigenvalues, in the diffusive domain, and obtain an expression for
its slope. Concluding remarks are made in section 10. 

\vskip1.5cm
\noindent
{\bf 2. Symmetries of the QCD Partition Function with Two Colors and 
Fundamental Fermions}
\vskip 0.5cm

The QCD partition function for two colors and fundamental fermions is 
special in the sense that it is invariant  under a larger symmetry
group than is the case for more colors.
This well-known fact for QCD with fundamental fermions relies on 
the pseudo-real nature of $SU_{\rm C}(2)$ \cite{Peskin,Shifman-three,SmilV}, 
and can be  easily  generalized  to the supersymmetric case \cite{zirnall}
relevant to the partially quenched QCD partition function. The discussion
in this section is mainly included for pedagogical reasons and part of it
was already given in reference \cite{Kyoto}.

The fermionic action is obtained by writing the determinant
in the partition
function as a Gaussian integral over the Grassmann fields $\phi$ and $\bar
\phi$. Using the chiral representation of the $\gamma$ matrices
with
\be
\gamma_\mu = \left ( \begin{array}
{cc} 0 & \hat \sigma_\mu  \\
                   \hat\sigma_\mu^+ & 0 \end{array} \right ),
\ee
and $\hat\sigma_\mu = (1, i\sigma_k)$
with $\sigma_k$ the Pauli $\sigma$-matrices,
the fermionic action can be written as
\be
S_F = \int d^4 x \sum_{f=1}^{N_f}
\left( \begin{array}{c} \bar \phi_R \\ \bar \phi_L \end{array} \right )
\left ( \begin{array}{cc} m_f &\hat\sigma_\mu (\partial_\mu + i A_\mu) \\
        \hat\sigma_\mu^+ (\partial_\mu + i A_\mu) & m_f \end{array} \right )
\left( \begin{array}{c}  \phi_R \\  \phi_L \end{array} \right ).\nonumber \\
 \ee
For $N_c = 2$, we have that $A_\mu^T = -\tau_2 A_\mu  \tau_2$ (with
$\tau_2 = \sigma_2$ in color space). Combining this
with the relation $\hat\sigma_\mu^* = \sigma_2 \hat\sigma_\mu \sigma_2$ we find
\be
\bar\phi_L^f \hat\sigma_\mu^+ (\partial_\mu + i A_\mu) \phi_R^f =
 \phi_R^f \sigma_2\tau_2 \hat\sigma_\mu (\partial_\mu + i A_\mu)
\sigma_2\tau_2 \bar \phi_L^f ,
\ee
where we have used that $\partial_\mu$ is anti-Hermitian and that the
fermion fields are anti-commuting Grassmann variables.
The fermionic action can thus be rewritten as
\be
S_F(m_f = 0) = \int d^4 x \sum_{f=1}^{N_f}
\left( \begin{array}{c} \bar \phi_R^f \\ \sigma_2 \tau_2 \phi_R^f \end{array} \right )
\left( \begin{array}{cc}\hat\sigma_\mu (\partial_\mu + i A_\mu)& 0 \\
        0& \hat\sigma_\mu (\partial_\mu + i A_\mu) \end{array} \right )
\left( \begin{array}{c}  \phi_L^f \\  \sigma_2\tau_2 \bar \phi_L^f \end{array} \right ).
\nonumber\\
 \ee
Obviously the symmetry group is $Gl(2N_f)$. If the number of left-handed
modes is not equal to the number of right-handed modes because of the
anomaly, an axial $U(1)$ is broken explicitly as for three or more
colors.

The mass term is given by
\be
S_m &=& \int d^4 x \sum_{f=1}^{N_f} \left [
\left( \begin{array}{c}  \phi_L^f\\  \sigma_2\tau_2 \bar \phi_L^f 
\end{array} \right )
\left ( \begin{array}{cc} 0& -\frac 12 m_f\sigma_2\tau_2  \\
        \frac 12 m_f\sigma_2\tau_2 & 0 \end{array} \right )
\left( \begin{array}{c}  \phi_L^f \\  \sigma_2\tau_2 \bar \phi_L^f 
\end{array} \right )
\right . \nonumber \\
&+& \left .
\left( \begin{array}{c} \bar \phi_R^f \\ \sigma_2\tau_2  \phi_R^f 
\end{array} \right )
\left ( \begin{array}{cc} 0& \frac 12 m_f\sigma_2\tau_2  \\
        -\frac 12 m_f\sigma_2\tau_2 & 0 \end{array} \right )
\left( \begin{array}{c} \bar \phi_R^f \\ \sigma_2\tau_2  \phi_R^f 
\end{array} \right )
\right ] .
\ee
Also in this case we expect a maximum spontaneous breaking of chiral
symmetry consistent  with the Vafa-Witten theorem \cite{Vafa-Witten}. 
This means that only the subgroup of $U(2N_f)$
that leaves both $\bar \phi_R \phi_R$ and $\bar \phi_L \phi_L$ invariant
remains unbroken. Therefore, only the subgroup  that leaves
\be
\left( \begin{array}{c}  \phi_L^f\\ \bar \sigma_2\tau_2 \bar \phi_L^f 
\end{array} \right )
\left ( \begin{array}{cc} 0& -\sigma_2\tau_2  \\
        \sigma_2\tau_2 & 0 \end{array} \right )
\left( \begin{array}{c}  \phi_L^f \\  \sigma_2\tau_2 \bar \phi_L^f 
\end{array} \right )
\ee
invariant remains unbroken ($\bar \phi_R \phi_R$ is invariant under
the same transformations). This is the symplectic group $Sp(2N_f)$.
The Goldstone manifold is thus given by the coset
$SU(2N_f)/Sp(2N_f)$. 

An analogous argument can be made for bosonic quarks. In that case
it is essential to allow only symmetry transformations that leave the
integrals in the partition function (\ref{pqQCDpf}) convergent.
In that case
one finds \cite{zirnall} that the chiral symmetry is broken according to
$Gl(2, R) \rightarrow O(2)$. 

The symmetries of
the partially quenched partition function for two colors with 
fundamental matter fields can be obtained along similar lines.
One easily convinces oneself that the symmetry of 
the Lagrangian is enhanced to $Gl(2N_f +2|2)$ from the usual group
$Gl(N_f+1|1) \times Gl(N_f+1|1)$  for more than two colors.  
For convergence reasons the symmetry group has to be restricted to 
a non-compact group $Gl(2, R)$ for the bosonic quarks.
This symmetry is broken spontaneously by the formation
of a chiral condensate in the boson-boson sector and the fermion-fermion
sector exactly as discussed above. The remaining symmetry group combines
into the ortho-symplectic graded Lie group $OSp(2N_f +2|2)$. The
Goldstone manifold is therefore given by \cite{zirnall}
the maximum Riemannian submanifold of the coset space
 $Gl(2N_f +2|2)/OSp(2N_f+2|2)$. This results in a integration domain for the
the fermion-fermion block given by the symmetric space $U(2N_f)/Sp(2N_f)$ and an
integration domain for the boson-boson block given by the symmetric
space $Gl(2,R)/O(2)$. 

An explicit parameterization of this Goldstone 
manifold is given by
\be
V^T G V,
\ee
where
\be
G= \left(
\begin{array}{ccc}
0 & 1_{\tiny N_f+1} & 0 \\
-1_{\tiny N_f+1} & 0 & 0 \\
0 & 0 & 1_{\tiny 2}
\end{array} \right)
\ee
and $V$ is parameterized  as
\be
V =\left(
\begin{array}{ccc}
U_{FF} & \alpha_1 & \alpha_2  \\
\beta_1 & a & b  \\
\beta_2 &c &d
\end{array} \right).
\ee  
Here, $\alpha_i$ and $\beta_i$ are Grassmann variables, $U_{FF}$ is an 
$(2N_f+2) \times(2N_f +2)$ unitary matrix and $a,\,b,\,c$ and $d$ are
real numbers.

\vskip 1.5cm
\noindent{\bf 3. Symmetries of the QCD Partition Function for Adjoint 
Fermions}
\vskip 0.5cm
For QCD with fermions in the adjoint representation the 
Dirac operator is given by
\be
\Dirac = \gamma_\mu\partial_\mu + f^{abc} \gamma_\mu A_{a\mu},
\ee
where the $f^{abc}$ denote the structure constants of the gauge group. 
In this case, the operator $\gamma_2\gamma_4 (\Dirac+m) $ is antisymmetric and
therefore the square root of the fermion determinant can 
be represented as a Grassmann integral \cite{HVeff} in the following way 
\be
&&{\det}^{1/2}( \gamma_2 \gamma_4(D+m)) = \nonumber \\&& \int d\psi_R d\psi_L
 \exp \left [
\left ( \begin{array}{c} \psi_R^f \\ \psi_L^f \end{array} \right ) 
\gamma_2 \gamma_4\left ( \begin{array}{cc} m & \sigma_\mu(\partial_\mu + 
f^{abc} \gamma A_{a\mu}) \\ \sigma_\mu^+(\partial_\mu +
f^{abc} \gamma A_{a\mu}) & m \end{array} \right )
\left ( \begin{array}{c} \psi_R^f \\ \psi_L^f \end{array}\right )\right ]
\nonumber \\
\ee
For simplicity we will consider only even $N_f$.
The flavor symmetry group
 is $SU(N_f)$ which acts on the Majorana fields  as
\be
\left ( \begin{array}{c} \psi_R \\ \psi_L \end{array} \right ) \rightarrow
\left ( \begin{array}{cc} U & 0 \\ 0 & U^{-1} \end{array} \right )
\left ( \begin{array}{c} \psi_R \\ \psi_L \end{array}\right ). 
\ee
This symmetry is broken by the chiral condensate which, with maximum 
breaking of chiral symmetry consistent with the Vafa-Witten theorem,
is given by (the color index is denoted by $i$ and the spinor index by
$\alpha$) 
\be
\Sigma_0 = \frac 2{N_f} \sum_{f=1}^{N_f}
\langle \epsilon_{\alpha \beta}(\psi^f_{R\,\alpha \, i}  
\psi^f_{R\, \beta \, i} +
\psi^f_{L\,\alpha \, i}\psi^f_{L\,\beta \, i}) \rangle,
\ee
and is invariant under an $O(N_f)$ subgroup of $SU(N_f)$. The Goldstone
manifold in this case is thus given by $SU(N_f)/O(N_f)$.

The square root of an anti-symmetric matrix cannot be represented as an
integral over real bosonic variables. Therefore, the above construction fails 
 for bosonic quarks.
However, it is possible to represent
the inverse determinant as an integral over complex variables. If we
use that
\be
\bar \phi_L \hat \sigma^+_\mu(\partial_\mu +
f^{abc} \gamma_\mu A_{a\mu}) \phi_R = (\sigma_2 \phi_R) \hat
\sigma_\mu(\partial_\mu +f^{abc} \gamma_\mu A_{a\mu})\sigma_2\bar\phi_L,
\ee
we observe that the flavor symmetry group of the bosonic quarks is doubled. 
In this case it turns out \cite{zirnall} that the chiral symmetry is
broken according to $U^*(2, R) \rightarrow Sp(2)$. 

The symmetry group of the partially quenched Lagrangian for matter
fields in the adjoint representation is $Gl(N_f+2|2)$. 
Because of convergence reasons only a $U^*(2, R)$ subgroup of the
boson-boson block of $Gl(N_f+2|2)$ is a symmetry of the partition function.
This symmetry is broken by the formation of a chiral condensate.
With maximum breaking of chiral symmetry consistent with the Vafa-Witten
theorem only an $OSp(2|N_f+2)$ subgroup remains unbroken.   
 The Goldstone manifold is then given by the maximum
Riemannian submanifold \cite{zirnall} 
of the coset space $Gl(N_f+2|2)/OSp(2|N_f+2)$
with fermion-fermion integration domain given by $U(N_f+2)/O(N_f+2)$
and boson-boson integration domain equal to $U^*(2,R)/Sp(2)$.

An explicit representation of this manifold is given by
\be
U= V^T G_A V,
\ee
where
\be
 G_A= \left(
\begin{array}{ccc}
 1_{\tiny N_f+2} &0 & 0 \\
0 & 0  & 1 \\
0 & -1 & 0
\end{array} \right)
\ee
and an explicit parameterization of $V$ is given by    
\be
V =\left(
\begin{array}{ccc}
U_{FF} & \alpha_1 & \alpha_2  \\
\beta_1 & a & b  \\
\beta_2 &c &d
\end{array} \right),
\ee
where $U_{FF}$ is an $(N_f+2) \times (N_f+2)$ unitary matrix, 
$\alpha_i$ and $\beta_i$ are 
Grassmann variables and $a,\, b, \, c$ and $d$ are real numbers.

\vskip1.5cm
\noindent{\bf 4. The Scalar Susceptibility in ChPT}
\vskip 0.5 cm

In this section we compute the scalar susceptibility according to 
standard ChPT with $N_f$ sea quarks and no valence quarks. 
The scalar susceptibility is defined by
\be
K^{ab} =-\int d^4x d^4 y {\rm Tr} t^a \langle  x | \frac 1{\Dirac+m} | 
y \rangle t^b \langle  y | \frac 1{\Dirac+m} | x\rangle.
\ee
For equal quark masses, the Dirac operator is flavor diagonal. Replacing 
the spatial integrals by a sum over the Dirac spectrum, we obtain

\be
K^{ab} = -\frac {\delta_{ab}}{2} \sum_k \frac 1{(i\lambda_k + m)^2}.
\ee

By introducing the isovector scalar source term $s^a \bar q t^a q$ 
the scalar susceptibility can
be expressed in terms of a derivative of the QCD partition function
(notice that the disconnected contribution vanishes in this case)
\be
K^{ab} = \partial_{s_a}\partial_{s_b} \log Z.
\ee

Alternatively, the low-energy limit of the scalar susceptibility can be obtained
from the effective chiral partition function. To lowest order in the momenta
and the quark masses this partition function is determined uniquely by the
pattern of chiral symmetry breaking \cite{GaL,foundations}
\be
\int_{U \in G/H} d U e^{ -\int d^4 x \left [\frac{F^2_\beta}{4} 
{\rm Tr}(\partial_\mu U \partial_\mu U^{-1}) 
- \frac{\Sigma_0}{2} {\rm Tr }(\hat{\cal M}_\beta (U+U^{-1}))\right ]}
\ee
The mass matrix also includes an isovector scalar source term. 
In all three cases, the matrix $U$ is parametrized as
\be
U = e^{i 2\pi_a T^a/F_\beta},
\ee
where the sum is over the generators $T^a$ of the coset (notice that
in this representation the mass matrix is diagonal for 
all three cases). The 
pion decay constant
$F_\beta$ is chosen such that we have the standard Gell-Mann-Oakes-Renner
relation for all three cases with the convention ${\rm Tr} T^a T^b =
\delta^{ab}/2$. 
For the three cases under consideration the appropriate parameters are given
in Table 1.
\renewcommand{\arraystretch}{1.3}
%\begin{center}
\begin{table}[h!]
\centering
\caption[]{\small The mass matrix and the pion decay constant for the
three different patterns of chiral symmetry breaking.}
\begin{tabular}{|c||c|c|c|}
\hline
$G/H$  & $\hat{\cal M}_\beta$ & $F_\beta$ & $\beta$\\
\hline
$SU(N_f)$ & $m {\bf 1}_{N_f}+ s_a t^a$ &$F$ &2 \\
$SU(2N_f)/Sp(2N_f)$ & $\left ( \begin{array}{cc} ( m
{\bf 1}_{N_f} + s_a t^a)/2 & 0 
                    \\0 & (m{\bf 1}_{N_f} + s_a t^{a\,T})/2  
\end{array}\right )$ 
&$ F/\sqrt 2$ &1 \\
$SU(N_f) /O(N_f)$  & $(m {\bf 1}_{N_f}+ s_a t^a)/2 $ & $F/\sqrt 2$& 4 \\
\hline
\end{tabular}
\end{table}
%\end{center}
\renewcommand{\arraystretch}{1.0}

To second order in the $\pi$ fields we find
\be
\frac 12(U + U^{-1}) = 1 - \frac 2{F^2_\beta} \pi_k \pi_l T^k T^l.
\ee
By differentiating the partition function with respect to the 
scalar source terms and calculating the loop integral we obtain
\be
K^{ab} &=& \frac {4 \Sigma^2}{F^4} {\rm Tr }(t^a T^k T^l)
{\rm Tr }(t^b T^m T^n)
\langle \int d^4 x d^4 y \pi_k(x) \pi_l(x) \pi_m(y) \pi_n(y) \rangle_{1-{\rm 
loop}}\nonumber \\ 
&=& \frac{\Sigma^2}{16\pi^2 F^4} 
{\rm Tr }(t^a \{T^k, T^l\}){\rm Tr }(t^b \{T^k, T^l\})\log (\Lambda/m).
\ee

The only nontrivial task is to calculate the sum
\be
Q^{(\beta)} = \frac 14 {\rm Tr }(t^a \{T^k, T^l\}){\rm Tr }(t^b \{T^k, T^l\})
\ee
Let us first consider the case $\beta = 2$. We already have seen that
 $K^{ab}$ is
diagonal in isospin space. For the generators $t^a$ and $t^b$ 
we conveniently choose
\be
t^a =t^b= \frac 12 (\delta_{ i1} \delta_{j1}-\delta_{ i2} \delta_{j2}).
\ee
The remaining diagonal generators are chosen as (consistent with the 
relation ${\rm Tr} \,t^k t^l = \frac 12\delta^{kl}$),
\be
{\rm diag}(t^n_D ) = (\underbrace{1, \cdots, 1}_n, -n, 
\underbrace{0, \cdots, 0}_{N_f-n -1})/\sqrt{2(n^2 +n)},
\label{diagonal}
\ee
whereas the off-diagonal generators are given by
\be
t_A^{kl} = i(\delta_{ik}\delta_{jl} - \delta_{jk}\delta_{il})/2,
\label{gena}
\ee
and
\be
t_S^{kl} = (\delta_{ik}\delta_{jl} + \delta_{jk}\delta_{il})/2,
\label{gens}
\ee
Notice that for $\beta = 2$ we do not distinguish between $T^a$ and $t^a$.
 For the sum $Q^{(2)} $ we thus obtain
\be
Q^{(2)}= \frac 14 \sum_{n=2}^{N_f-1}  \frac 1{n(n+1)} + \frac 1{16}(N_f-2)
= \frac 1{16}\frac{N_f^2 - 4}{N_f},
\label{sumgen}
\ee
where the first term in  this equation is due to the 
diagonal generators 
and the second term of this equation is
due to the off-diagonal generators.

As usual the case of $\beta =2 $ is the simplest case. The next simplest case 
is $\beta = 4$ with coset space given by $SU(N_f)/O(N_f)$. 
In the sum (\ref{sumgen}) the same diagonal
generators enter as in the case of $\beta = 2$. However, only the symmetric 
off-diagonal generators contribute to this sum in this case, so
that the contribution from the off-diagonal generators is exactly half
of what it was for $\beta = 2$. This results in 
\be
Q^{(4)}= \frac 14 \sum_{n=2}^{N_f-1}  \frac 1{n(n+1)} + \frac 1{32}(N_f-2)
=\frac 1{32}\frac{(N_f+4)(N_f-2)}{N_f}.
\ee

Finally we consider the case $\beta =1$. In this case the generators of
the coset have the structure \cite{Peskin}
\be
T^k = \left ( \begin{array}{cc} A & B \\ B^\dagger & A^T \end{array} \right),
\ee 
where  $A$ 
is a Hermitian traceless matrix and $B^T = - B$
(both are $N_f \times N_f$ matrices).
The sum over the generators
is again calculated by distinguishing the diagonal generators from
the off-diagonal generators. The diagonal generators are given by
\be
T^a_D = \frac 1{\sqrt 2} \left ( \begin{array}{cc} t^a_D & 0 \\ 0 & t^{a}_D
\end{array} \right ),
\ee
where the $t^a_D$ are defined in (\ref{diagonal}).
In order to satisfy the symmetry requirements, the off-diagonal generators
have four nonzero entries in this case. 
Adding up all contributions we find
\be
Q^{(1)}= 1{8}\frac {N_f -2}{N_f}+\frac 1{8} (N_f -2)= 
\frac 1{8} \frac{(N_f-2)(N_f+1)}{N_f}
\ee
%%%% Notice that the generator from the source term is normalized differently
%%%% from the other generators.
where the first term 
is the contribution from the diagonal generators and
the second term is the contribution from the off-diagonal generators.

Our final result for $K^{ab}$ can be summarized as
\be
K^{ab} = -\frac{\Sigma^2}{32 \pi^2 F^4} 
\frac {(N_f-2)(N_f+\beta)}{\beta N_f} \log 
(m/\Lambda).
\ee
With the same assumption for the $\lambda$ and $m$ dependence of the
spectral density as in \cite{SmilgaStern} we find that the slope
of the Dirac spectrum at the origin is given by
\be
\rho'(0) = \frac{\Sigma^2}{16 \pi^2 F^4}
\frac {(N_f-2)(N_f+\beta)}{\beta N_f}.
\ee
The slope vanishes for 
$N_f = 2$ for all three values of $\beta$. 
This is in agreement with instanton simulations \cite{Vinst} performed for
the cases of $\beta= 1 $ and $\beta = 2$.

\vskip1.5cm
\noindent
{\bf 5. Partially Quenched Chiral Perturbation Theory}
\vskip 0.5cm

The physics of QCD at low energies is completely determined
by the spontaneous symmetry breaking pattern described above
\cite{GaL,foundations,GL,LS}. In general, the effective Lagrangian
of partially quenched QCD is
based on the spontaneous chiral symmetry breaking pattern
of the graded Lie-groups $G\rightarrow H$. The Goldstone fields are
given by
\be
U = \exp (i 2 \Phi/F_\beta),
\label{uphi}
\ee
where $F_\beta$ is the pion decay constant and $\Phi = \phi_a T^a$ with
$T^a$ the generators of the  maximum Riemannian submanifold of $G/H$
(notice that in this representation the mass matrix is diagonal for all three 
cases).
We use the normalization ${\rm Str} T^a T^b = \frac 12 \delta^{ab}$.
To lowest order in the momenta and the quark masses, 
the Euclidean effective Lagrangian is obtained 
in the very same way as in Chiral Perturbation Theory \cite{GaL,foundations,
qChPT,pqChPT}. This results in
\be
{\cal L}_{\rm eff}=\frac{F^2_\beta}{4} {\rm Str}(\partial_\mu U 
\partial_\mu U^{-1}) 
- \frac{\Sigma_0}{2} {\rm Str}(\hat{\cal M}_\beta (U+ U^{-1}))
+{m_0^2} \Phi_0^2
+{\alpha} \partial_\mu \Phi_0 \partial_\mu \Phi_0 \label{Leff}.
\ee
The last two terms in (\ref{Leff}) represent the mass and kinetic terms of the 
super-$\eta'$ flavor-singlet field $\Phi_0={\rm Str}(\Phi)$. They are 
introduced for the same symmetry reasons as in the three-color 
case \cite{qChPT,pqChPT}. Below we are only interested in the case
that $m_0 \sim \Lambda_{QCD}$. In that case the term proportional to $\alpha$
is sub-leading, and it will be ignored from now on.
The effective partition function is given by
\be
Z_{\rm eff} = \int_{U\in G/H} d U e^{-\int d^4 x {\cal L}_{\rm eff}},
\label{effpart}
\ee 
where the coset, the mass matrix and the pion decay constant for the
three values of $\beta$ are given in the table below. 

\begin{table}[h]
\centering
\caption[]{\small The mass matrix and the pion decay constant for the
three different patterns of chiral symmetry breaking for the partially
quenched effective partition function.}
\renewcommand{\arraystretch}{1.3}
\begin{tabular}{|c||c|c|c|}
\hline
$G/H$  & $\hat{\cal M}_\beta$ & $F_\beta$ & $\beta$\\
\hline
$Gl(N_f+1|1)$ & $ \hat{\cal M}= \left(
                 \begin{array}{cc}
                 M_2   & 0\\
                 0 &   m_v
                 \end{array} \right) $ &$F$ &2 \\
$Gl(2N_f+2|2)/OSp(2N_f+2|2)$ & $\hat{\cal M}= \left(
                 \begin{array}{cccc}
                 M_1/2 &0 & 0 & 0\\
                 0 & M_1/2  & 0 & 0\\
                 0 & 0 & m_v/2 & 0 \\
                 0 & 0 & 0 & m_v/2
                 \end{array} \right) $ &$F/\sqrt 2$ &1 \\
$Gl(N_f+2|2)/OSp(2|N_f+2)$  & $\hat{\cal M}= \left(
                 \begin{array}{cccc}
                 M_4/2  & 0 & 0\\
                 0 &  m_v/2 & 0 \\
                 0 &  0 & m_v/2
                 \end{array} \right) $ &$F/\sqrt 2$ &4 \\
\hline
\end{tabular}
\end{table}
\renewcommand{\arraystretch}{1.3}
The mass matrix in the fermion-fermion block of the mass matrix 
${\cal M}_\beta$ is given by
\be
M_2&=&{\rm diag} (\underbrace{m_s, \dots, m_s}_{N_f},m_v+J),\\
M_1&=&{\rm diag} (\underbrace{m_s, \dots, m_s}_{N_f},m_v+J),\\
M_4&=&{\rm diag} (\underbrace{m_s, \dots, m_s}_{N_f},m_v+J,m_v+J).
\ee

For the zero momentum component of the flavor 
singlet mass term, the same physical interpretation as in the three-color case 
holds \cite{OTV}:
\be
\langle \nu^2\rangle = \frac {F^2_\beta m_0^2 V}2. 
\label{topol}
\ee 
The factors in front of the first two terms in the effective Lagrangian 
(\ref{Leff}) have been chosen such that the Goldstone fields are
normalized in the usual way. The masses of the Goldstone bosons
are given by the usual Gell-Mann-Oakes-Renner relation
\be
M^2_{ij}=(m_i+m_j) \; \Sigma_0/F^2. \label{GOR}
\ee
with  quark condensate in the chiral limit denoted by $\Sigma_0$.
 
In the case of chiral perturbation theory for QCD it is possible to
distinguish a domain where the partition function factorizes in a piece
dominated by the zero momentum modes and a factor involving the nonzero
momentum modes \cite{GL,LS}. The same is true for the partially quenched
chiral Lagrangian.  For valence quark masses below the Thouless energy 
given by \cite{Vplb,Osbornprl,OTV}
\be
m_c \sim \frac {F^2}{\Sigma_0 L^2},
\ee
the  fluctuations of the zero modes in the 
the effective partition function dominate the fluctuations of the nonzero
momentum modes. Then the integral over the zero momentum modes 
factorizes from the partition function and the calculation of
$\Sigma(m_v)$ is reduced to a group integral.
This integral has been calculated explicitly for $\beta= 2$
with the result that $\Sigma(m_v)$ is in
complete agreement with chiral Random Matrix Theory.

\vskip1.5cm
\noindent
{\bf 6. Calculation of $\Sigma(m_v)$ for $\beta = 1$}
\vskip 0.5cm

From a computational point of view, two-color pqChPT  is not very
different from three-color pqChPT \cite{pqChPT}. To perform a calculation to
one-loop order one has to construct the meson propagator. The effective
Lagrangian discussed in previous section is based on the symmetric
super-space $Gl(2 N_f+2|2)/OSp(2 N_f+2|2)$.  
An explicit representation of $\Phi$ in (\ref{uphi}) in terms of
a $(2 N_f+4) \times (2 N_f+4)$ matrix field is given by
\be
\Phi=\left(
\begin{array}{cccc}
\phi & \psi & \eta_v/2 & \eta_w/2 \\
\psi^\dagger & \phi^T & \chi_v/2 & \chi_w/2 \\
\chi_v/2 &  -\eta_v/2 & i\tilde{\phi}_v/\sqrt 2 & i\tilde{\phi}_{vw}/2 \\
\chi_w/2 & -\eta_w/2 & i\tilde{\phi}_{vw}/2 &  i\tilde{\phi}_w/\sqrt 2
\end{array} \right),
\ee
where the $(N_f+1) \times (N_f+1)$ ordinary matrices 
$\phi=\phi^\dagger$ and 
$\psi=-\psi^T$ contain the ordinary mesons made of quark and anti-quarks,
the $N_f+1$-vectors $\chi_v$, $\chi_w$, $\eta_v$ and $\eta_w$ 
represent the fermionic mesons 
consisting of a ghost quark and an ordinary anti-quark, and finally the
real fields $\tilde{\phi}_v$,  $\tilde{\phi}_w$ and  $\tilde{\phi}_{vw}$
are the mesons with two ghost quarks. 
With the proper normalization of the fields the Hermitian matrix field 
$\phi$  is given by
\be
\phi = \left (\begin{array}{ccc} 
\phi_{1,1}/2 & \cdots & \phi_{1,N_f+1}/2\sqrt 2 \\ 
    \vdots &\ddots & \vdots \\
 \phi_{1,N_f+1}^*/2\sqrt 2 & \cdots & \phi_{N_f+1,N_f+1}/2
 \end{array} \right ),  
\ee
and the anti-symmetric matrix $\psi$ is given by 
\be
\psi = \left (\begin{array}{cccc} 
0 &  \psi_{1,2}/2\sqrt 2&\cdots & \psi_{1,N_f+1}/2\sqrt 2 \\ 
   -\psi_{1,2}/2\sqrt 2&\ddots & &\vdots  \\ 
    \vdots & & &\\
  &&&  \psi_{N_f,N_f+1}/2\sqrt 2 \\ 
   -\psi_{1,N_f+1}/2\sqrt 2&\cdots & -\psi_{N_f,N_f+1}/2\sqrt 2
& 0 
  \end{array} \right ).  
\ee
To second order in the pion fields there is no mixing between 
the  off-diagonal Goldstone modes.  
However, because of the term ${\rm Str}^2 \Phi$, the situation is more
complicated for the diagonal Goldstone modes. The propagator of these
modes is obtained by inverting the quadratic form in the effective 
Lagrangian obtained by expanding the $U$ fields up to second order.
The mixing matrix of the diagonal mesons is given by
\be
M^{(1)} = \left ( \begin{array}{ccccc} 
2 & \cdots & 2& -i \sqrt 2 & -i \sqrt 2 \\
\vdots &    &\vdots &\vdots & \vdots \\
2 & \cdots & 2& -i \sqrt 2 & -i \sqrt 2 \\
-i\sqrt 2 & \cdots & -i\sqrt 2 & -1 & -1 \\ 
-i\sqrt 2 & \cdots & -i\sqrt 2 & -1 & -1  \end{array}\right )
\ee
The propagator of the diagonal mesons is thus given by
\be
G^{(1)}(p^2) = (\Gamma + m^2_0 M^{(1)})^{-1},
\ee
where $\Gamma$ is diagonal with
\be
\Gamma_{ii} = \left \{ \begin{array}{ccc}
 p^2 + M_{ss}^2 \quad &{\rm for}& \quad 1\le i \le N_f,\\
 p^2 + M_{vv}^2 \quad &{\rm for}& \quad N_f+1\le i \le N_f+3, \end{array} 
\right  . 
\ee
The matrix $\Gamma+M^{(1)}$ can be inverted by expansion in a power series
in $\Gamma^{-1} M^{(1)}$ and using the relation that
\be
(\Gamma^{-1} M^{(1)})^2 = 2 N_f (p^2+M_{ss}^2)^{-1}\Gamma^{-1} M^{(1)}.
\ee
By re-summing the geometric series we obtain
\be
G^{(1)}(p^2) = \frac 1\Gamma -
\frac 1{(p^2 + M_{ss}^2)^{-1} + 2N_f m^2_0} m^2_0 \frac 
1\Gamma M^{(1)} \frac 1\Gamma.
\ee
Below, we only need this  propagator at the origin
for $1 \le i \le N_f+1$
which is given by
\be
G_{ii}^{(1)}=\frac 1V \sum_p \left [\frac{1}{p^2+M_{ii}^2}-\frac{2 m^2_0\; 
(p^2+M^2_{ss})}
{ (p^2+M^2_{ii})^2 \; (p^2+M^2_{ss}+2N_f m^2_0 )} \right ]. 
\label{prop}
\ee

The quenched limit is obtained in the limit of infinite sea quark masses
at fixed values of $m_0$ and $m_v$. Standard ChPT power counting rules are 
recovered when the super-$\eta'$ decouples, 
that is when the singlet mass $m_0$ is sent to infinity.
It is easy to check that in this case the propagator (\ref{prop}) in the
sea quark sector is just the standard propagator in a quark basis with the
singlet channel projected out. Finally, if both the valence and sea quark 
masses are set equal, we recover the standard result for 
two-color ChPT.

The valence quark mass dependence of the chiral condensate is computed as in 
the three color case \cite{GolLeung,OTV}. Since we are mainly 
interested in the non-analytic valence quark mass dependence of the chiral
condensate, we will perform a one-loop computation, and we will ignore 
the analytic corrections coming from the $O(p^4)$ effective Lagrangian.

The usual tadpole diagrams  have to be evaluated. At finite volume, the 
pion propagator at the origin is 
\be
\Delta(M^2)= \frac  1V \sum_p   \;
\frac{1}{p^2+M^2},
\ee
where the sum is over the momenta in the box. In the thermodynamic limit
the sum can be replaced by an integral resulting in (in four Euclidean
dimensions)
\be
\Delta(M^2) =\frac{1 }{16 \pi^2} M^2 {\rm ln}
\frac{M^2}{\Lambda^2}, 
\ee 
where $\Lambda$ is the cutoff.

It is now straightforward to calculate 
the valence quark mass dependence of the chiral condensate. Collecting the
different contributions we find
\be
\Sigma^{(1)}(m_v)&=&
\Sigma_0\left [ 1 - \frac 1{F^2} \left \{ 2N_f \Delta(M_{vs}^2)
+ G_{vv}^{(1)} - 2\Delta(M_{vv}^2) \right \} \right ] 
\nonumber \\
&=&
\Sigma_0 \; \Bigg[ 1- \frac{1}{N_f F^2} \Bigg\{ 2 N_f^2 
\Delta(M^2_{vs}) 
-(N_f+1) \Delta(M^2_{vv}) \nonumber \\ &&+(M^2_{ss}-M^2_{vv}) 
\partial_{M^2_{vv}}\Delta(M^2_{vv}) \Bigg\} \Bigg]. \label{sigma}
\ee
The first term in the first equation 
is due to the off-diagonal mesons consisting out of a valence quark and a sea
quark,
the second term is due to the diagonal meson consisting out of two valence
quarks, and the last term is due to mesons consisting out of a bosonic 
and a fermionic quarks. As usual there is an additional minus sign for
fermionic meson loops.

\vskip1.5cm
\newpage
\noindent
{\bf 7. Calculation of $\Sigma(m_v)$ for $\beta = 4$}
\vskip 0.5cm
For fermions in the adjoint representation the Goldstone manifold 
is determined
by the pattern of chiral symmetry breaking according to
\be
Gl(N_f+2|2) \rightarrow OSp(2|N_f+2).
\ee
The Goldstone manifold is given by the maximum Riemannian submanifold
of $Gl(N_f+2|2) /OSp(2|N_f+2)$. 
The effective Lagrangian has exactly the same structure as for $\beta =1$,
but in this case an explicit representation of the Goldstone modes is
given by
\be
\Phi=\left(
\begin{array}{ccccc}
A_{1,1}/\sqrt 2 &\cdots& A_{1,N_f+2}/2 & \alpha_1/2 &\beta_1/2\\
\vdots & \ddots     &\vdots      & \vdots   & \vdots\\
A_{1,N_f+2}/2& \cdots &A_{N_f+2,N_f+2}/\sqrt 2 &\alpha_{N_f+2}/2 & 
\beta_{N_f+2}/2\\
\beta_1/2&\cdots &\beta_{N_f+2}/2 & i\phi_1/ 2 & 0\\
-\alpha_1/2 & \cdots& -\alpha_{N_f+2}/2& 0 & i\phi_1/2 \end{array}\right ) .
\ee
Here, $A$ is a real symmetric matrix, $\phi_1$ is a real variable and the
$\alpha_k$ and $\beta_k$ are Grassmann variables.

In order to do a one-loop 
calculation we need an explicit expression
for the pion propagator. Because of the singlet mass term the inversion
of the quadratic form involving the diagonal mesons is nontrivial.
The propagator of the diagonal mesons is given by
\be
G^{(4)}(p^2) = ( \Gamma + m^2_0 M^{(4)})^{-1},
\label{prop4}
\ee
where the mixing matrix is given by
\be
M^{(4)} = \left ( \begin{array}{cccc}
1 & \cdots & 1& -i \sqrt 2  \\
\vdots &    &\vdots &\vdots \\
1 & \cdots & 1& -i \sqrt 2 \\
-i\sqrt 2 & \cdots & -i\sqrt 2&-2  \\
\end{array}\right )
\ee
and $\Gamma$ is diagonal with
\be
\Gamma_{ii} = \left \{ \begin{array}{ccc}
 p^2 + M_{ss}^2 \quad &{\rm for}& \quad 1\le i \le N_f,\\
 p^2 + M_{vv}^2 \quad &{\rm for}& \quad N_f+1\le i \le N_f+3. \end{array} 
\right  . 
\ee

The inverse in (\ref{prop4}) can be  calculated simply
by expanding the expression as a geometric
series in $\Gamma^{-1} M^{(4)}$. If we use the relation
\be
(\Gamma^{-1} M^{(4)})^2 = {N_f} \frac 1{p^2 +M_{ss}^2}\Gamma^{-1} M^{(4)}
\ee
the geometric series can be resummed as
\be
G^{(4)}(p^2) = \frac 1{\Gamma} - \frac 1{1+(p^2 + M_{ss}^2)^{-1} N_f m^2_0}
m^2_0\frac 1\Gamma M^{(4)} \frac 1\Gamma.
\ee
Below we only need the first $N_f+2$ diagonal elements of the propagator
matrix at the origin. They are given by
\be
G_{ii}^{(4)} =\frac 1V \sum_p \left [
 \frac 1{p^2 + M_{ii}^2} - \frac{m_0^2(p^2+M_{ss}^2)}
{(p^2+M_{ii}^2)^2(p^2 +M_{ss}^2 + N_f m_0^2)} \right ].
\ee
The calculation of the valence quark mass dependence of the
chiral condensate proceeds in exactly the same way as for $\beta = 1$.
As final answer we find
\be
\Sigma^{(4)}(m_v) = \Sigma_0\left [1 -\frac 1{2F^2} 
\left \{ N_f\Delta(M_{sv}^2)
+ 2G_{vv}^{(4)}  - \Delta(M_{vv}^2)\right \} \right ].
\ee

\vskip 1.5cm
\noindent {\bf 8. Valence Quark Mass Dependence of the Chiral Condensate and
the Dyson Index}
\vskip 0.5cm
Before studying the valence quark mass dependence of the chiral condensate 
in five different limiting cases we write $\Sigma(m_v) $ as a single formula
valid for all three different values of $\beta$. 

The result for the valence quark mass dependence of the chiral condensate
for gauge fields in the fundamental representation and three or more
colors was obtained  previously  \cite{GolLeung,OTV}. 
In our present notation the result can be rewritten as
\be
\Sigma^{(2)}(m_v) = \Sigma_0\left [1 -\frac 1{F^2} 
\left \{ N_f \Delta(M_{sv}^2)
+ G_{vv}^{(2)}  - \Delta(M_{vv}^2)\right \} \right ],
\ee
where $G_{vv}^{(2)} = G_{vv}^{(4)}$.
With Dyson index $\beta$ of the Dirac operator equal to $\beta=1$ for QCD
with two colors and fermions in the fundamental representation, $\beta = 2$
for three or more colors and fermions in the fundamental representation, 
and $\beta =4$ for two or more colors in the adjoint representation, the result
for the three different cases considered above can be written as
\be
\Sigma^{(\beta)}(m_v) = \Sigma_0\left [1 -\frac 1{F^2} 
\left \{ \frac {2N_f}{\beta} \Delta(M_{sv}^2)
+ G_{vv}^{(\beta)}   - \frac 2 \beta\Delta(M_{vv}^2)\right \} \right ].
\label{general}
\ee
The diagonal propagator at the origin is given by
\be
G_{vv}^{(\beta)} = \frac 1V \sum_p \left [
\frac 1{p^2 + M_{vv}^2} - \frac{(1+\delta_{\beta,1})
m_0^2(p^2+M_{ss}^2)}
{(p^2+M_{ii}^2)(p^2 +M_{ss}^2 + N_f (1+\delta_{\beta,1})m_0^2)} \right ]
.
\ee
Notice that this propagator becomes independent of $\beta$ in the limit
 $m_0 \rightarrow \infty$ when the usual ChPT counting rules prevail.
Previously, a similar unified description was obtained for sum-rules for 
the inverse eigenvalues of the Dirac operator \cite{V,SmilV}.
Below we will discuss five limiting cases for 
$\Sigma(m_v)$ obtained in  (\ref{general}).

\noindent i) In the limit $m_v=m_s\ll m_0$ we should recover 
the mass dependence of the chiral condensate in standard chiral perturbation
theory. In this case with $M_{vv}=M_{ss}=M$, the diagonal propagator reduces to
\be
G_{vv}^{(\beta)} = \frac{N_f -1}{N_f} \Delta(M^2),
\ee
and the expression for $\Sigma(m_v)$ can be written as
\be
\Sigma^{(\beta)}(m_v) = \Sigma_0\left [1 -\frac 1{F^2} 
\frac {(N_f-1)(1+2N_f/\beta) }{ N_f} \Delta(M^2) \right ].
\ee
We observe that for $N_f =1$ the result is independent of the Dyson index
$\beta$. Indeed, in this case the only axial symmetry is broken explicitly 
by the anomaly for all three values of $\beta$. The 
factor $(N_f-1)(1+2N_f/\beta) $ exactly counts the number of Goldstone 
bosons which is equal to the number of generators
of $SU(2N_f)/Sp(2N_f)$ for $\beta =1$, the number of generators of
$SU(N_f)$ for $\beta = 2$, and the number of generators of $SU(N_f)/O(N_f)$
for $\beta = 4$.

\noindent ii) Next we consider the case that both $m_v \ll m_0$
and $m_s \ll m_0$. In this case
the diagonal propagator is given by
\be
G_{vv}^{(\beta)} =  \frac{N_f -1}{N_f} \Delta(M_{vv}^2) 
-\frac 1{N_f} {M_{vv}^2}\partial_{M_{vv}^2} \Delta(M_{vv}^2).
\ee
This results in the valence quark mass dependence
\be
\Sigma(m_v) = \Sigma_0\left [1 -
\frac 1{F^2}\left \{ \frac {2N_f}\beta \Delta(M^2_{vs}) 
+(\frac{N_f -1}{N_f} -\frac 2\beta )
\Delta(M_{vv}^2) + \frac 1{N_f} (M_{ss}^2 - M_{vv}^2) 
\partial_{M_{vv}^2} \Delta(M_{vv}^2) \right \} \right ].\nonumber \\
\ee
In the thermodynamic limit, the singular part of $\Sigma(M_v)$ is then given 
by
\be
\Sigma(m_v)=\Sigma_0 \; \Bigg[ 1- \frac{\Sigma_0}{8 \pi^2 N_f F^4} 
\Bigg\{ \frac{N_f^2}\beta (m_v+m_s) \log \frac{m_v+m_s}{2 \mu}
+\Big( m_s+(N_f(1- \frac 2\beta) -2) m_v \Big) 
\log \frac{m_v}{\mu} \Bigg\} \Bigg], \nonumber \\
\label{sigmaT}
\ee
where $\mu=\Lambda^2 F^2/2 \Sigma_0$. 
In the limit of $m_s \rightarrow 0$ this result simplifies to
\be
\Sigma(m_v)=\Sigma_0 \; \Bigg[ 1- \frac{\Sigma_0}{8 \pi^2 N_f\beta F^4} 
 (N_f-2)(N_f +\beta) m_v \log \frac{m_v}{\mu}
\Bigg]\label{sigmaTmsz}.
\ee

\noindent iii)
In the range $\lambda_{min} \ll m_v \ll 1/L^2 \Lambda_{\rm QCD}\ll m_0$, 
and $m_s \ll m_v$ the 
propagator at the origin (\ref{prop}) is dominated by the zero-momentum mode
\cite{GL,HasLeu,OTV,DOTV} resulting in
\be
\Delta(M^2_{vv}) \sim \frac 1{M_{vv}^2 V}.
\ee
The valence quark mass dependence of the chiral condensate in the limit 
$m_s \ll m_v$ then reduces to
\be
\Sigma_v \sim   \Sigma_0\Bigg[1 - (\frac {2N_f}{\beta}
+ \frac 12 - \frac 1\beta )\frac 1{  m_vV \Sigma_0}\Bigg].
\label{sigmaV}
\ee

\noindent iv) 
Next we consider the quenched limit which is 
obtained for $m_s \gg m_v$ and $m_s \gg m_0$. In this case the
diagonal propagator at the origin is given by
\be
G_{vv}^{(\beta)} = \frac 1V \sum_p \left [
\frac 1{p^2 +M_{vv}^2} - \frac {(1+\delta_{\beta,1})
m_0^2}{(p^2 + M_{vv}^2)^2} \right ] .
\ee
This results in the valence quark mass dependence
\be
\Sigma(m_v) &=& \Sigma_0\left [1 -\frac 1{F^2} 
\left \{ (1-2/\beta)\Delta(M_{vv}^2) + 
{(1+\delta_{\beta,1})m_0^2}
\partial_{M_{vv}^2}
\Delta(M_{vv}^2)\right \} \right ]\nonumber \\
&\sim &
\Sigma_0\left [1 -\frac 1{16\pi^2F^2}  {(1+\delta_{\beta,1})m_0^2}
\log \frac{m_v}\mu 
 \right ] + O(m_v).
\label{quenched}
\ee
where  we have only quoted the infrared singular terms.
We find that the chiral condensate diverges for $m_v \rightarrow 0$.

In the domain $m_v \ll m_s$ the expression (\ref{sigmaT}) is identical
to the quenched result (\ref{quenched}) provided 
$(1+\delta_{\beta,1})m_0^2= M^2_{ss}/N_f$. This
is in accordance with the Ward identity relating the sea quark mass and
the topological susceptibility $\langle \nu^2\rangle = F^2_\beta 
M_{ss}^2 V/2 N_f(1+\delta_{\beta,1})$ \cite{Shifman}, 
and the relation (\ref{topol}).

\noindent v) In lattice QCD with staggered fermions the Lagrangian does not
possess the axial $U(1)$ symmetry at finite lattice spacing (although it
should be recovered in the continuum limit). A detailed
discussion of quenched lattice simulations for the valence quark mass
dependence of the chiral condensate in this case was given in
\cite{chiral-logs}.
The effective partition function is
then given by (\ref{effpart}) with an effective Lagrangian 
without singlet terms.
The diagonal propagator at the origin is thus given by
\be
G_{vv} = \frac 1V \sum_p  \frac 1{p^2+M_{vv}^2} 
\ee
in the cases i), ii) and iv). In the quenched case, the term $\sim m_0^2$
is absent and the $O(m_v)$ terms have to be taken into account. We then find
\be
\Sigma(m_v) = \Sigma_0\left [ 1 - \frac {(1-2/\beta)\Sigma_0}
{8\pi^2 F^4}m_v\log \frac {m_v}\mu \right ].
\ee
For $\beta = 2$ the coefficient of the chiral logarithm vanishes
and higher order corrections become important \cite{Rakow}.
\newpage
\vskip1.5cm
\noindent
{\bf 9. Spectral Density in the Ergodic and Diffusive Domains}
\vskip 0.5cm

In the ergodic regime, for valence quark masses $m_v \ll F^2/\Sigma_0 L^2$
the microscopic spectral density is given by chRMT \cite{Vplb}, whereas for
$\Lambda_{\rm QCD} \gg 
m_v \gg \lambda_{\rm min}\sim 1/L^4 \Sigma_0$ the one-loop calculation
of $\Sigma(m_v) $ is valid. In this section 
we calculate the spectral densities in each of these
 domains.  We first show that in the ergodic domain
the one-loop result for $\Sigma(m_v)$ coincides with the chRMT result.
 
The QCD partition function for two colors and fundamental fermions
belongs to the chGOE $(\beta = 1)$ universality class  \cite{V}. 
In a sector of
topological charge $|\nu|$ and in the chiral limit, 
the result reads \cite{VGOE,TiloHab}:
\be
\rho_s^{\rm chGOE}(u)&=&\frac{u}{2} [J^2_{2 N_f+|\nu|}(u)-J_{2 N_f+|\nu|+1}(u) 
J_{2 N_f+|\nu|-1}(u)] \nonumber \\
&&+\frac{1}{2} J_{2 N_f+|\nu|}(u) \; [1-\int_0^u dt J_{2 N_f+|\nu|}(t)].
\label{micro1}
\ee

The QCD Dirac operator with adjoint fermions
belongs to  the chGSE $(\beta = 4)$ universality class. 
In this case the microscopic 
spectral density is given by \cite{nagao-forrester,TiloHab}
\be
\rho_s^{\rm chGSE}(u)&=&{u} [J^2_{N_f+2|\nu|}(2u)-J_{ N_f+2|\nu|+1}(2u) 
J_{N_f+2|\nu|-1}(2u)] \nonumber \\
&&-\frac 12 J_{2 N_f+2|\nu|}(2u) \int_0^{2u} dt J_{ N_f+2|\nu|}(t).
\label{micro4}
\ee
For completeness we also give the microscopic spectral density for the 
chGUE $(\beta=2)$ \cite{VZ,Vinst} which describes the Dirac operator of 
QCD with three or more colors
and fundamental fermions,
\be
\rho_s^{\rm chGUE}(u)&=&\frac{u}{2} [J^2_{N_f+|\nu|}(u)-J_{N_f+|\nu|+1}(u) 
J_{ N_f+|\nu|-1}(u)]. \nonumber \\
\label{micro2}
\ee

The valence quark mass dependence of the chiral condensate in terms 
of the microscopic variable $x=m_v V \Sigma_0$ is obtained via the relation
\be
\frac{\Sigma(x)}{\Sigma_0} = \int_0^\infty du \frac{2x}{u^2+x^2} \rho_s(u)
+\frac{|\nu|}{x}, \label{sigmaR}
\ee
where the last term is the contribution from the exactly 
zero eigenvalues of the Dirac operator \cite{Dampart}.
In the case of the chGUE this integral can be perform analytically resulting
in
\be 
\frac{\Sigma(x)}{\Sigma_0} &=& x [ I_{N_f+|\nu|}(x)K_{N_f+|\nu|}(x)
+I_{N_f+|\nu|+1}(x)K_{N_f+|\nu|-1}(x)] +\frac{|\nu|}{x}\nonumber \\
&=&1- \frac{N_f +|\nu|} x +\frac {|\nu|}x +\cdots= 1 -\frac{N_f}x +\cdots.
\ee
The asymptotic result for $x \gg 1$ for the chGOE and the chGSE can be
easily derived using this result and the asymptotic result of the remaining
terms. The leading $1/x$ correction can be obtained by expanding before
integration. The only additional integral that is required is
\be
\int_0^\infty J_\mu(x) =1.
\ee
The asymptotic result in terms of the Dyson index is given by
\be 
\frac{\Sigma_v(x)}{\Sigma_0} 
 = 1 -\frac{4N_f +\beta-2}{2\beta x},
\ee
which coincides with the result (\ref{sigmaV}) 
obtained from chiral perturbation theory.
Notice, that the leading order asymptotic correction is independent of $\nu$.
This is no longer true for higher order corrections and a summation over
all $\nu$ 
is required to compare with results obtained for fixed value
of $\theta$ (which is zero in our case) \cite{poulsum}. 

In the diffusive domain, for $1/L^2 \Lambda_{\rm QCD} \ll m_v \ll \lambda_{\rm 
QCD}$ the valence quark mass dependence of the chiral condensate is
given by the one-loop result (\ref{sigmaT}). The spectral density is given 
by the discontinuity of $\Sigma(m_v)$ 
across the imaginary axis 
($\rho(\lambda)/V = \left .{\rm Disc}\right |_{m_v = 
i\lambda}\Sigma(m_v)/2\pi$) \cite{OTV,DOTV}.
As final result we then find for the spectral density of the Dirac operator
with $N_f$ flavors with quark masses $m_s \ll \lambda \ll \Lambda_{\rm QCD}$
\be 
\frac{\rho (\lambda)}V
&=&\frac{\Sigma_0}{\pi} \Bigg[ 1+\frac{\Sigma_0}{16 \pi^2 N_f F^4 }
\Bigg\{ \frac{2 N^2_f}{\beta} |\lambda| {\rm Arctg} \frac{|\lambda|}{m_s}
-(N_f(\frac 2\beta -1)+2) \pi |\lambda|  \\
&& \hspace{4cm} -\frac{N^2_f}\beta m_s \log
\frac{\lambda^2+m^2_s}{\mu^2}-2 m_s \log \frac{|\lambda|}{\mu} \Bigg\}
\Bigg] .\nonumber
\label{rhomlam}
\ee
The spectral density has a logarithmic infrared divergence when $m_s \neq 0$. 
In the limit $m_s \ll \lambda$, the expression (\ref{rhomlam}) reduces to:
\be 
\frac{ \rho(\lambda)}V &=&\frac{\Sigma_0}{\pi}
\Bigg[ 1+\frac{ (N_f-2) (N_f+\beta) \Sigma_0}{16 \pi \beta N_f F^4 }  
|\lambda|\Bigg].
\label{rhoQCD} 
\ee
The result for the slope for $\beta =2$ was first obtained in
\cite{SmilgaStern}. Remarkably, the slope 
vanishes for $N_f=2$ for three values of $\beta$. This is in agreement
with instanton liquid simulations  of the QCD Dirac
spectrum \cite{Vinst}.

In the quenched case, the spectral density is found to be:
\be
\frac{ \rho^Q(\lambda) }V = \frac {\Sigma_0}\pi\left [ 1 - 
\frac {(1+\delta_{\beta,1})m_0^2}{16\pi^2 F^2} \log \frac {|\lambda|}\mu
\right ].
\ee
We find a logarithmic divergence of the spectral density for all three values
of $\beta$. This result is consistent with quenched 
 instanton liquid and lattice QCD simulations \cite{Osbornnpb,TeperHart}.  
on liquid simulations and
lattice QCD simulations \cite{Osbornnpb,TeperHart}. For staggered fermions
away from the continuum limit this term has to be modified according
to the discussion given in v) of the previous section. For the spectral
density we then find
\be
\frac{ \rho^Q(\lambda) }V =\frac {\Sigma_0}\pi\left [ 1 -
\frac{\Sigma_0}{16\pi F^4} (1- \frac 2\beta) |\lambda|
\right ].
\ee

\vskip1.5cm
\noindent
{\bf 10. Conclusions}
\vskip0.5cm
We have shown that the infrared limit of the QCD Dirac spectrum is 
completely determined
by the pattern of chiral symmetry breaking. The spectral density 
follows from the
low energy effective partition function of
the partially quenched 
QCD partition function which in addition to the usual quarks contains
valence quarks and their bosonic superpartners. In this
paper we have analyzed the case of two colors with fundamental 
fermions and the case of fermions in the
adjoint representation.
In both cases the effective Lagrangian is based on a Riemannian super-manifold
originating from chiral symmetry breaking to an ortho-symplectic graded 
Lie Group. As is the case for QCD with three or more colors with
fundamental fermions the Goldstone 
manifold is characterized by a symbiosis of
 compact and noncompact degrees of freedom.

Our   one-loop calculation of the valence quark mass 
dependence of the chiral condensate for these two chiral symmetry breaking
patterns completes an earlier calculation for QCD with three colors
and fundamental fermions \cite{OTV}. Its discontinuity results in an 
explicit expression for the spectral density of the Dirac operator valid
for $\lambda \ll \Lambda_{\rm QCD}$.
In all three cases we can distinguish two different domains in 
the Dirac spectrum
which are separated by the Thouless energy. For valence quark masses 
below the Thouless energy the kinetic term decouples from the effective
Lagrangian and the  the one-loop result for
the valence quark mass dependence of the chiral condensate is in complete
agreement with chiral Random Matrix Theory. Agreement with chiral 
Random Matrix Theory to all orders has only been shown for $\beta =2$
by the means of an exact calculation of the superintegrals. It would be
interesting to perform this calculation for the mathematically much more
complicated cases of $\beta=1$ and $\beta=4$ as well.
For valence quark masses beyond the Thouless
energy the kinetic term in the partially quenched effective partition
function has to be taken into account. The results for
the spectral density in this domain are consistent with the results 
for the scalar susceptibility obtained with standard ChPT.
In particular we have obtained an analytical expression for the slope of 
the Dirac spectrum. Remarkably, the slope vanishes for two massless flavors 
in all three cases.

\vskip 1.5cm
\noindent
{\bf Acknowledgements}
\vskip 0.5cm

This work was partially supported by the US DOE grant
DE-FG-88ER40388. One of us (D.T.) was supported in part by
``Holderbank''-Stiftung.
A. Altland, P. Damgaard, M. G\"ockeler, J. Osborn, 
P. Rakow, A. Smilga, M. Stephanov, 
T. Wettig and M. Zirnbauer are acknowledged for useful discussions.

\end{document}